# Broad-Spectral-Range Sustainability and Controllable Excitation of Hyperbolic Phonon Polaritons in α-MoO$_3$

Weikang Dong, Ruishi Qi, Tiansheng Liu, Yi Li, Ning Li, Ze Hua, Zirui Gao, Shuyuan Zhang, Kaihui Liu, Jiandong Guo, and Peng Gao*

W. K. Dong, R. S. Qi, N. Li, Z. Hua, Prof. P. Gao,
International Center for Quantum Materials, and Electron Microscopy Laboratory, School of Physics, Peking University, Beijing, 100871, China.
E-mail: p-gao@pku.edu.cn

Prof. T. S. Liu
School of Environment and Safety Engineering, North University of China, Taiyuan 030051, China.

Y. Li, S. Y. Zhang, Prof J. D. Guo
Beijing National Laboratory for Condensed Matter Physics and Institute of Physics, Chinese Academy of Sciences, Beijing, 100190, China.

Z. R. Gao
Beijing National Laboratory for Molecular Sciences, College of Chemistry and Molecular Engineering, Peking University, Beijing, China.

Prof. K. H. Liu
State Key Laboratory for Mesoscopic Physics, School of Physics, Peking University, Beijing 100871, China.

Prof. K. H. Liu, Prof. P. Gao
Collaborative Innovation Center of Quantum Matter, Beijing 100871, China.

**ABSTRACT**

Hyperbolic phonon polaritons (HPhPs) in orthorhombic-phase molybdenum trioxide (α-MoO$_3$) show in-plane hyperbolicity, great wavelength compression and ultra-long lifetime, therefore holding great potential in nanophotonic applications. However, its polaritonic response in the far-infrared (FIR) range has long remained unexplored due to challenges in experimental characterization. Here, using monochromated electron energy loss spectroscopy (EELS) in a scanning transmission electron microscope (STEM), we probe HPhPs in α-MoO$_3$ in both mid-infrared (MIR) and FIR frequencies and correlate their behaviors with microstructures and orientations. We find that low-structural symmetry leads to various phonon modes and multiple Reststrahlen bands (RBs) over a broad spectral range (over 70 meV) and in different directions (55–63 meV and 119–125 meV along *b* axis, 68–106 meV along *c* axis, 101-121 meV along *a* axis). These HPhPs can be selectively excited by controlling the direction of swift electrons. These findings provide new opportunities in nanophotonic and optoelectronic applications such as directed light propagation, hyperlenses and heat transfer.

Surface phonon polaritons (SPhPs) have great potential in achieving low-loss nanophotonic applications such as waveguiding, superlensing, enhanced optical forcing, enhanced energy transfer and sensing[1-4]. In the frequency band between transverse optical (TO) and longitudinal optical (LO) frequencies (termed RB), the real part of permittivity becomes negative and sustains SPhPs. In hyperbolic media such as hexagonal boron nitride (h-BN)[4-7] and α-MoO$_3$, the permittivity has opposite signs along different axes, giving rise to even more exotic physics that provides avenues to novel polaritonic applications such as subdiffraction imaging[6] and super-Planckian thermal emission[4]. Several recent works have demonstrated that α-MoO$_3$, a natural van der Waals material with full anisotropy along three Cartesian axes, supports HPhPs in the MIR range with many desirable properties such as in-plane hyperbolicity, deep-subwavelength light confinement and topological transitions, making it arguably one of the most promising polaritonic materials[8-12].

In fact, α-MoO$_3$ has multiple RBs ranging from MIR to FIR, with different kinds of dispersions in each of them. However, previous studies haven't fully revealed the intriguing properties of HPhPs in α-MoO$_3$ since they only involved two RBs in MIR. Technically, the difficulty is that the commonly employed method for visualizing SPhPs - scattering-type scanning near-field optical microscopy (s-SNOM) – is generally limited to the MIR (typically >100 meV) and the terahertz range (<10 meV)[13, 14], leaving a gap covering FIR due to the lack of continuous-wave laser sources. A very recent work[10] combined s-SNOM with photo-induced force microscopy (PiFM) to extend the accessible spectral range a little bit, but the energy region below ~100 meV still remained unexplored.

Recent advances in monochromated STEM-EELS have enabled spectroscopy measurements in an extremely wide continuous spectral range (typically 30 meV–3 keV) with sub-10-meV energy resolution and subnanometer spatial resolution, providing a powerful tool for SPhP visualization in individual nanostructures[15-24]. This motivates the present study using monochromated STEM-EELS to probe the full polaritonic response of α-MoO$_3$.

In this paper, we study the effects of anisotropy on HPhP launching and propagation in α-MoO$_3$ over a broad spectral range. We find that low structural symmetry leads to various phonon modes and anisotropic permittivities in three orthogonal crystal axes, and the overlap of four RBs associated with eight phonon modes makes HPhP sustainable in a broad spectral range (over 70 meV). Our measurements reveal an additional RB in FIR along *b* axis and HPhPs within it, which have never been experimentally reported or theoretically predicted before, and also provide direct experimental observation of HPhPs in another RB along *c* axis that were not fully explored due to technical difficulties. By controlling the direction of the electron beam, these HPhPs can be selectively excited. These observations can be well reproduced by numerical simulations. Such a broad spectral range sustainability and controllable excitation of HPhPs in a single material may find potential applications in nanophotonics and optoelectronics.

Shown in **Figure 1**a is the atomic structure of α-MoO$_3$, in which each Mo atom is surrounded by six O atoms and a unit layer is composed of two layers of such octahedrons. Notice that in each octahedron (inset), the six O atoms are not equivalent. Such a complex structure, along with longitudinal-transverse splitting effects[25], gives rise to a great variety of phonon modes in the MIR to FIR region, as confirmed by the Raman spectrum in Figure 1b. In our context, four pairs of TO and LO phonons ranging from 55 meV to 125 meV are relevant: one pair vibrating along *a*, two pairs along *b* and one pair along *c*. From phonon frequencies, permittivities along three axes can be calculated by the multiple Lorentz oscillator model, as presented in Figure 1c (Note S1). The shaded regions denote RBs along corresponding directions, the overlap of which makes HPhPs sustainable in a broad spectral range with a width of over 70 meV.

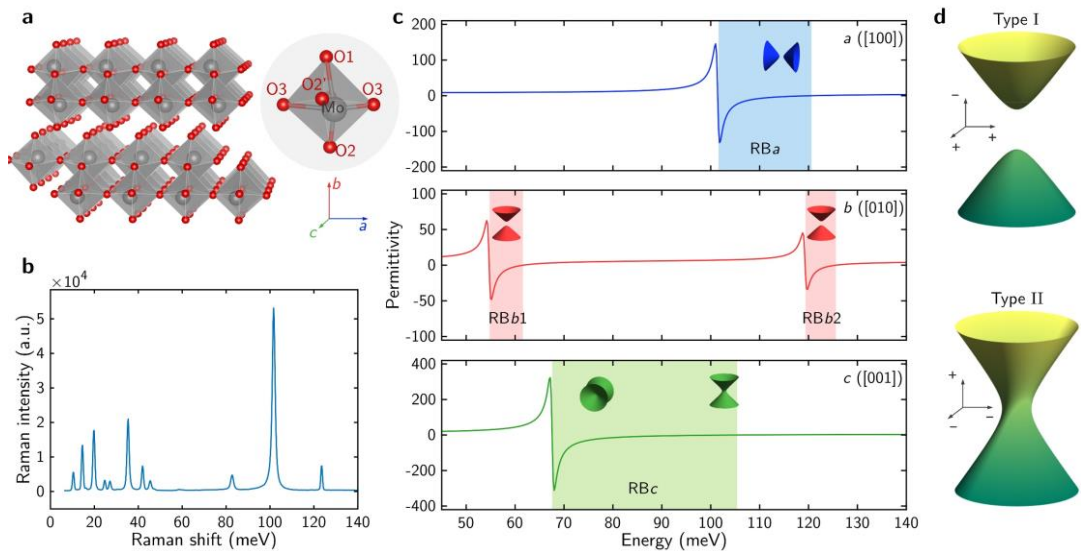

**Figure 1.** Phonon modes and anisotropic permittivity of α-MoO$_3$. a) Atomic structure of α-MoO$_3$. Inset shows an octahedron composed of one Mo atom and six O atoms. b) Raman spectrum of α-MoO$_3$, showing multiple phonon peaks. c) Anisotropic permittivity along three principal axes. Top, middle and bottom panels show permittivity along *a*, *b* and *c* axes, respectively. Light shadows mark four RBs. d) Schematic of isofrequency surface in wavevector space. Upper and lower panels correspond to type I and type II dispersions, respectively. This is also shown as small insets in (c).

Note that at different energy, the signs of permittivities along three directions can achieve many different combinations. As illustrated in Figure 1d, when two of them are positive while the other one is negative, isofrequency surface in the momentum space forms a hyperboloid of two sheets (Type I dispersion); when two are negative while the other one is positive, it forms a hyperboloid of one sheet (Type II dispersion)[8]. According to the isofrequency surface, the propagation of HPhPs is highly anisotropic. For both types, wavevectors along axes with positive permittivity have no intersection with the hyperboloid, and thus cannot be supported. At different energy region α-MoO$_3$ shows different types and directions of hyperbolicity, as schematically shown in insets in Figure 1c.

**Figure 2**a illustrates the experimental configuration of the STEM-EELS technique. Kiloelectronvolt electrons are focused to atom-wide and then interact with the sample, launching SPhPs upon losing a tiny fraction of their kinetic energy (tens of meV in tens of keV). Through the energy loss of low-angle scattered electrons, information of vibration properties can be extracted. Meanwhile, high-angle scattered electrons are collected to simultaneously retrieve information about sample morphology or atomic structure. Figure 2b presents a high-angle annular dark field (HAADF) image of an α-MoO$_3$ sample. To ensure excitation of phonon modes in all three directions, the sample is placed in an arbitrary orientation (i.e., the beam is not parallel to any of three principal axes; see Note S2 and Figure S1 in Supporting Information for detailed discussion). Figure 2c shows a typical EEL spectrum acquired with the electron beam located near the sample but without intersecting it (aloof

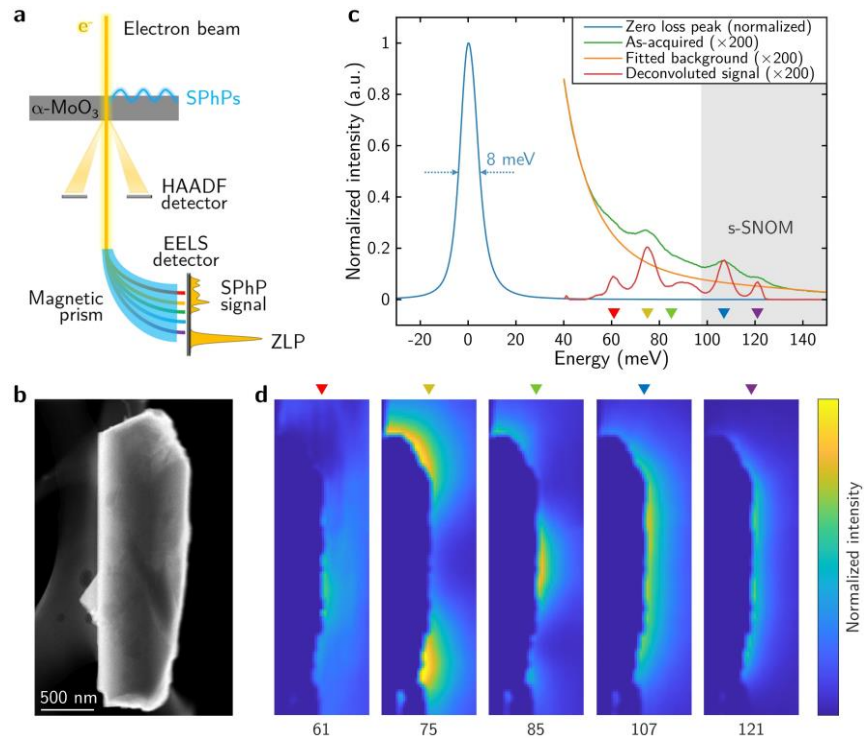

**Figure 2.** Probing SPhPs using STEM–EELS. a) Schematic of STEM-EELS experimental setup. b) HAADF image of an α-MoO$_3$ sample placed in an arbitrary orientation with respect to the electron beam. c) Typical EEL spectrum taken in aloof geometry. Blue, EEL spectrum normalized by its maximum value, showing a sharp ZLP with a width of 8 meV (FWHM). Green, magnified spectrum to show the energy loss area. Yellow, fitted background which is subtracted from the spectrum to extract vibrational signal. Red, deconvoluted signal, showing 5 major energy loss peaks. Light gray shadow marks the energy region commonly investigated by optical methods in previous works. d) Energy-filtered EELS maps at several energy (numbers below the maps in meV). The energy is also marked on the top by small triangular markers that appear in (c) with the same color.

geometry). The blue curve shows a typical zero loss peak (ZLP) with a full width at half maximum (FWHM) of only 8 meV, demonstrating the monochromaticity of the electrons. On the right side of the ZLP, we found resonant peaks due to coupling between the electrons and lattice vibrations. After background subtraction and deconvolution, we simultaneously obtain 5 major resonant peaks (red curve) between 50-130 meV. As a comparison, the shaded region in Figure 2c marks the accessible range of optical methods, which only covers less than a half of the signal.

With the beam rastering around the sample, we obtain two-dimensional (2D) EELS maps at specific energy. Quantitatively, these EELS maps represent the electromagnetic density of states (DOS) projected along the electron beam[26], which is a local property of SPhPs. Two energy loss peaks at 61 meV and 121 meV lie in RB*b*1 and RB*b*2; peaks at 75 meV and 85 meV in RB*c* are resulted from vibrations along *c* direction; and the peak at 107 meV can be attributed to HPhPs in RB*a*. All these peaks lie in the corresponding RBs shown in Figure 1c, thus confirming our permittivity model. Note that RB*b*1 has not been reported or predicted before to the best of our knowledge due to its low energy. Note S3 and Figure S2 in Supporting Information further confirm the observed signal in this RB is truly from phonon polaritons in α-$MoO_3$ instead of artifacts introduced during data processing. In RB*c*, previous optical studies only covered a narrow spectral band at higher energy, leaving the lower region unexplored[8-10]. In contrast, the swift electrons can simultaneously excite all these SPhPs in all four RBs and allow us to map them from the EEL spectra. In RB*c* (75 and 85 meV) we observe nodal patterns with alternate distribution of nodes and anti-nodes along the sample, which can be interpreted as interference fringes of SPhPs. Note that although the energy and intensity of polariton eigenmodes in a finite sample are dependent on sample size and geometry, the observed features in this study are due to the orientation-dependent excitation of HPhPs rather than purely due to shape effects (Note S4 in the Supporting Information).

When the electron beam trajectory is parallel to one of the three axes, we observe selective excitation of HPhPs. **Figure 3**a shows a HAADF image of a natural α-$MoO_3$ nanoribbon with its long axis along *c* and short axis along *a*, as indicated by white arrows[27]. The electron beam is travelling perpendicular to the figure plane (parallel to *b* axis). Shown in Figure 3b–3d are 2D EELS maps in RB*c*, RB*a* and RB*b*2 respectively, taken near the edge of the nanoribbon (dashed box in Figure 3a). In RB*c*, HPhPs launched by the electron beam propagate towards the terminal, and are then reflected back. The interference between incident and reflected HPhPs leads to nodal patterns. With higher energy, the polariton wavelength decreases, so the interference maximum moves toward the ribbon terminal. This is very similar to propagating SPhPs in an isotropic nanowire, which has been extensively explained elsewhere[15]. Figure 3e is a line profile with the electron beam scanning along *c* axis. To corroborate our experimental results, we performed finite-difference time-domain (FDTD) simulation[28] (Figure 3f), which agrees nicely with the experiment. As a reference, Figure S4 shows corresponding boundary element method (BEM) simulation[29] for the isotropic case, taking permittivities along *a*, *b* and *c* respectively. The propagating mode in RB*c* shows an obvious energy shift as a manifestation of their dispersion relation, which is similar to the propagating mode in the isotropic simulation (Figure S4c). However, in RB*a* such kind of energy shift associated with the interference of propagating HPhPs is absent compared with the isotropic case, which can be confirmed by the homogeneity of energy loss signal along the ribbon in Figure 3c. Such behaviors can be explained by their hyperbolic dispersion resulting from the strong anisotropy. As mentioned above in Figure 1d, when only $\varepsilon_a$ is negative while the other two are positive, the isofrequency surface of the Type I dispersion is a hyperboloid with two sheets that has no intersection with the *bc* plane, forbidding wavevectors along *c*. Their propagation along *a* (short edge of the ribbon) results in interference fringes parallel to the long axis of the ribbon, so the signal in Figure 3d is generally homogeneous when scanning along *c*. The propagation of HPhPs in α-MoO$_3$ is naturally in-plane anisotropic, which is consistent with previous s-SNOM measurements in higher frequency region[9] and can be utilized to achieve novel nanophotonic applications, such as controllable thermal emission[30, 31] and directed light transmission without the need of complicated nano-patterning[10, 32].

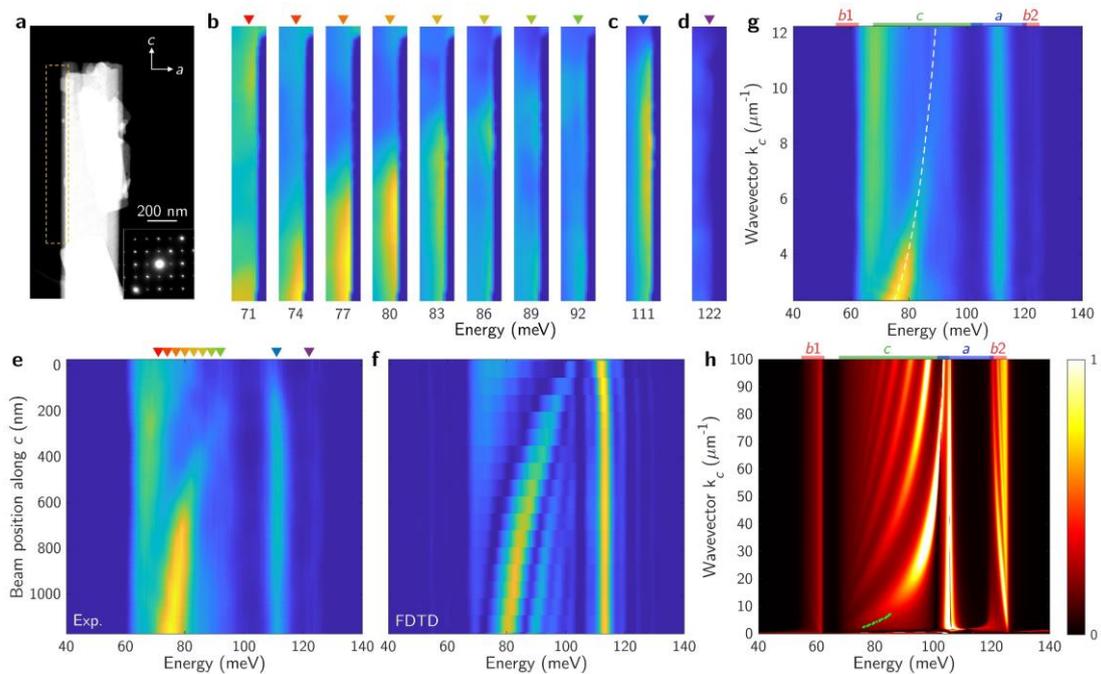

**Figure 3.** Selective excitation of in-plane HPhPs in a natural α-MoO3 nanoribbon. a) HAADF image of a ~100-nm-thick nanoribbon. Inset is an electron diffraction pattern taken near the edge of the nanoribbon. b-d) Energy-filtered EELS maps in three RBs, taken in the yellow dashed box in (a). e) Experimental EELS line profile along *c* direction. The origin of the vertical axis is located at the ribbon terminal. f) Simulated EELS line profile using FDTD. g) Dispersion relation of HPhPs transformed from the real-space experimental data shown in (e). Dashed white line is a guide to the eye. h) Calculated imaginary part of complex reflectivity, showing the dispersion relation of HPhPs along *c*. Green dots are extracted from our EELS measurement. In (g) and (h), lines in blue, red and green on the top denote RBs.

Another observation one should notice is that the energy loss intensity in RB$b$1 and RB$b$2 almost vanished (Figure 3d), indicating that the excitation is selective (Note S2). This can be understood by the fact that the electric field of the beam is predominately perpendicular to the beam trajectory. The ability to selectively excite HPhPs using electron beams may be useful for novel applications.

For a quantitative analysis, we extract the dispersion of HPhPs in RB$c$ from real-space data. Previous analytical calculation has shown that the reflection phase shift at the flake boundary quickly approaches $\pi/4$ as the wavelength suppression goes stronger[33, 34]. This means the distance between the first interference maximum and the ribbon terminal corresponds to $\frac{7\pi}{8k_c}$, where $k_c$ is the polariton wavevector along $c$ direction. Then the dispersion relation of HPhPs in RB$c$ can be extracted by transforming the real-space data into the momentum space, as presented in Figure 3g. Good agreement is found between our measurement and the calculated momentum-dependent complex reflectivity shown in Figure 3h (Note S6). Note that at similar frequency the free-space wavelength of photons is on the order of 10 µm, while the polariton wavelength is much smaller (e.g., ~50 times smaller at 90 meV), allowing deep-subwavelength light confinement.

To investigate anisotropic dispersion in other directions, we used focused ion beam (FIB) etching to obtain flakes with square shape, whose two lateral lengths are approximately equal and thus will not introduce additional asymmetry in boundary conditions to the system. **Figure 4**a shows a HAADF image of a fabricated square flake, whose lateral length is about 2.5 µm with a thickness of ~1 µm along $a$ (parallel to the beam trajectory). Figure 4b and 4c show line profiles with the beam scanning along *b* and *c* directions, respectively, and corresponding simulation is shown in Figure S6 (Note S7). Figure 4d shows two-dimensional EELS maps of several polariton eigenmodes. For comparison, simulated EELS maps of corresponding eigenmodes using FDTD are presented in Figure 4e below the experimental maps. Although in Figure 4d EELS signals inside the flake are all blocked due to insufficient penetration depth of the electron beam, from EELS data recorded in aloof geometry we can still find reasonable agreement between the simulation and the experiment (see Note S7 and Figure S5 in Supporting Information for a more quantitative comparison). Unlike those modes in isotropic nanocubes[16], polariton modes here show strong anisotropy in their spatial distribution. At 60 meV (in RB*b*1) and 120 meV (RB*b*2), the interference fringes of the dipolar mode are predominately parallel to *c* direction, indicating its propagation is anisotropic and prefers *b* direction. This also holds true for multipolar modes at 63 meV and 123 meV. Similarly, at 89 meV (in RB*c*) its propagation prefers *c* direction. We also observed a corner mode at 74 meV, whose DOS is concentrated near the corner of the flake. In the RB parallel to the electron trajectory (at 110 meV in RB*a*), there is no distinguished difference between *b* and *c*, and the excitation efficiency is much lower, which is consistent with our result previously presented in Figure 3.

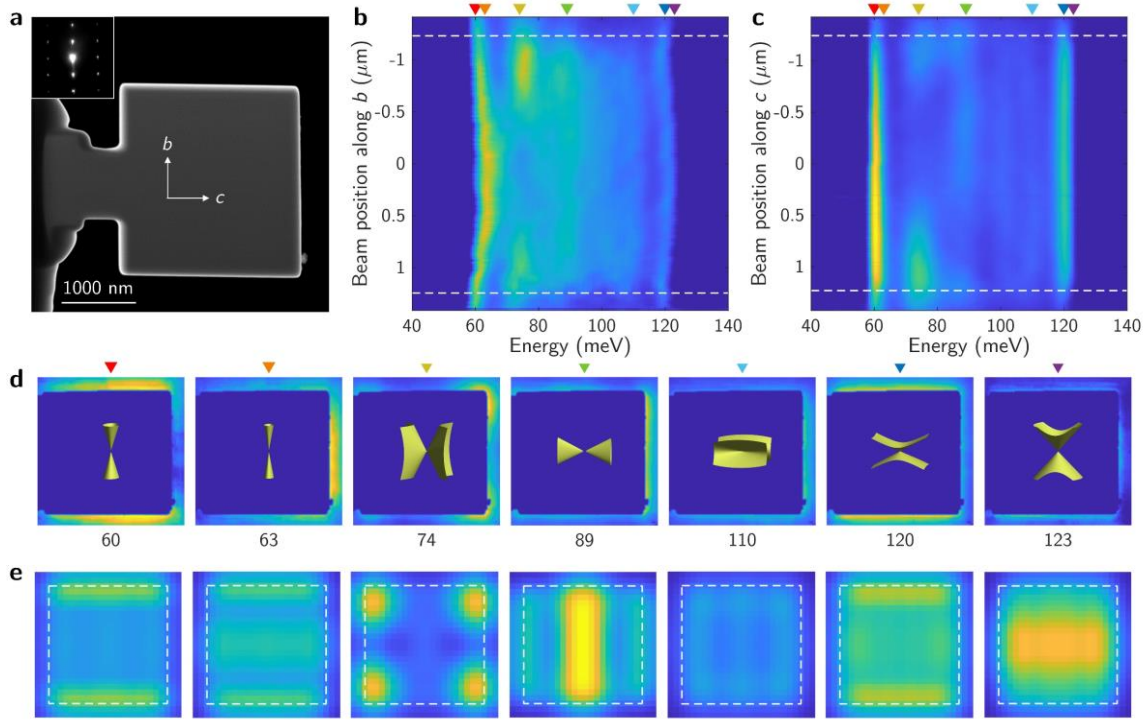

**Figure 4.** Hyperbolic phonon polaritons in *bc* plane. a) HAADF image of a MoO$_3$ flake fabricated by FIB. Inset shows the corresponding electron diffraction pattern, confirming its crystal orientation. b,c) Line profiles with the beam scanning along *b* and *c* directions. Horizontal dashed lines mark the edge of the flake. d) Energy-filtered EELS maps at selected energy marked by triangular markers of corresponding color. Insets illustrate isofrequency surfaces calculated by Fresnel equation. e) Simulated EELS maps of several polariton eigenmodes corresponding to (d). White dashed boxes denote the boundary of the flake.

Similarly, we also fabricated α-MoO$_3$ squares with two edges along *a* and *b* directions. The EELS data is consistent with our conclusion above, as shown in Figure S7 (Note S8).

We have used an electron probe to launch and detect HPhPs in α-MoO$_3$ over a broad energy range. Our EELS signal, with the spacing between two antinodes equals half wavelength, is similar to the so-called "edge-reflected" polaritons observed in optical measurements[7, 10, 35]. While this technique does obviate the difficulty in FIR laser sources, its limitations are also obvious. The energy resolution in this technique, limited by monochromaticity of the kiloelectronvolt electrons, is still sub-10 meV, losing fine structures of the spectra. Moreover, short penetration depth of the beam precludes data acquisition with intersecting geometry for thick samples. These problems may be addressed with the fast development of electron microscopy in the future.

In summary, we use STEM-EELS to map HPhPs in α-MoO$_3$ at MIR to FIR wavelengths. We find that α-MoO$_3$ is capable to support HPhPs in a remarkably wide spectral region (over 70 meV) with hyperbolicity along various directions, which allows deep-subwavelength concentration of the electromagnetic field and controllable propagation directions. We also observe selective excitation of HPhPs by simple control of the electron beam incident direction. We anticipate that in other polar crystals with low symmetry, such as recently reported orthorhombic-phase vanadium pentoxide[36], their polaritonic response would be similar, i.e., various types of HPhPs can also be supported over a broad spectral range and can be selectively excited. These findings provide useful information of anisotropic RB and HPhPs in materials with low structure symmetry for future potential applications in nanophotonics and polaritonics.

**Experimental Section**

*Sample synthesis:* Bulk α-MoO$_3$ crystals were synthesized by the vapor transport method in a high-temperature horizontal tubular furnace with a quartz tube (4.5 cm inner diameter and 120 cm in length). The heater surrounding the tube was 20 cm in length at the longitudinal center, where the metal source (molybdenum metal foil with the purity of 99.95%, Alfa Aesar) was located. The Mo source was heated at 800 °C in O$_2$ with the flow rate of 10 sccm for 3 hours. After cooling down to room temperature, transparent, centimeter-scale α-MoO$_3$ flakes were obtained on the inner wall of the tube away from the center by about 10 cm at both sides.

*FIB etching:* A FIB system (FEI STRATA DB235) was employed to prepare α-MoO$_3$ flakes along *bc* and *ab* plane. The crystal orientation was first determined through edges of the as-grown bulk crystals under the scanning electron microscope incorporated in the FIB system. It was confirmed later by electron diffraction patterns in our STEM measurements. In order to minimize possible sample damage during the etching process, a 20 nm carbon film was first deposited on the sample by a precision etching coating system (GATAN Model 682), and another Pt layer with about 0.5 μm thickness was then deposited by Ga ion beam induced deposition. A micrometer-scale rectangular flake was first etched off from the bulk crystal using a relatively large beam current (~ 400 pA) and was stuck to a copper TEM grid. It was then finely etched to the intended size using a small beam current (~ 50 pA).

*EELS data acquisition:* The EELS data was acquired using a Nion Ultra STEM$^{TM}$ 200 microscope equipped with both a monochromator and aberration correctors, with 60-kV beam energy, 15-mrad beam convergence semi-angle and ~ 5-pA beam current. Collection semi-angle was 24.4 mrad with a 1 mm spectrometer entrance aperture. Typical energy resolution (FWHM of ZLP) was 7 - 9 meV. Typical acquisition time was 400 ms per pixel in order to achieve a satisfactory signal-to-noise ratio. All EELS data was acquired using Nion Swift software.

Due to strong scattering of electrons from α-MoO$_3$, it's estimated that for every 15 nm inside α-MoO$_3$ the beam intensity decays to 1/e (Figure S8). Such a fast decay makes EEL spectra inside the sample highly noisy and vibration information is hard to extract. Therefore, we choose relatively large thick samples (typically > 100 nm) and aloof mode to extract information of HPhPs.

*EELS data processing:* To correct the energy shift of the ZLP, all the spectra were first aligned by cross correlation and then normalized by the integrated intensity of the unsaturated part of the ZLP to exclude influence of beam current fluctuation. After alignment and normalization, we applied block-matching and 3D filtering (BM3D) algorithm to remove Gaussian noise[37, 38]. Each energy channel is a 2D data containing two spatial dimensions that can be individually denoised. Based on high-frequency elements in the Fourier domain, noise level of each channel is individually estimated and then used as an input parameter in the BM3D denoising algorithm. We find that this process not only improves the signal-to-noise ratio, but also benefits the fitting in the following background-subtraction process. Figure S9a and S9b illustrate its effects on data quality.

From the denoised data, vibrational signal is extracted by subtracting a curve fitted to the background. The fitting is performed over two energy windows, approximately 20 meV wide both in front of and after the loss area. The function used to fit the background is $\exp[P_3(x)]$, where $P_3(x)$ is a cubic polynomial with its coefficients as fitting parameters. All the peaks we get after background subtraction have been individually scrutinized to make sure that they are all real signals rather than artifacts of background subtraction. The major criterion is that all signals should decay while the beam moves away from the sample.

Finally, in order to remove the broadening effect of the signal caused by limited monochromaticity of the electron beam, we applied Lucy-Richardson deconvolution algorithm to recover unblurred spectra and improve the energy resolution[39, 40]. The ZLP was used as the point spread function for 10 – 20 iterations to exclude the broadening and extract the signal (Figure S9c).

*BEM and FDTD simulation:* The overall picture for EELS simulation can be described as below. The electron beam produces an electric field, which interacts with the α-$MoO_3$ nanostructure characterized by a diagonalized complex permittivity tensor. The dielectric induces another field, which forces back on the electrons. The energy loss of the electron is therefore proportional to the integral of the $z$-component of the induced field where $z$ denotes the direction of the beam velocity. Following this picture, EELS can be simulated by solving Maxwell's equations for the induced field.

EELS simulation using FDTD has been demonstrated in literature[28]. This is done by transforming the electron beam source into a series of dipole sources. After solving Maxwell's equations and monitoring the induced electronic field, the energy loss of the electrons is then computed by integrating the $z$-component of the induced electric field along the trajectory.

BEM simulation of EELS using MNPBEM Toolbox has been well established and its high accuracy in simulating isotropic dielectrics proved[15, 16]. It works by solving Maxwell's equations via BEM, which only meshes up the boundary of the dielectric instead of volume and therefore only need a relatively small cost of memory and computation time, but requires the isotropy and homogeneity of the dielectric.

Typically, the surface was divided into about 1000 boundary elements in our simulation. Detailed description of methodology can be found in literatures[29, 41, 42].

**Supporting Information**

Supporting Information is available from the Wiley Online Library or from the author.

**Acknowledgments**

W.D. and R.Q. contributed equally to this work. The work was supported by the National Key R&D Program of China (2016YFA0300804), the National Natural Science Foundation of China (11974023 and 51672007), the Key R&D Program of Guangdong Province (2018B030327001, 2018B010109009, 2019B010931001), the National Equipment Program of China (ZDYZ2015-1), Bureau of Industry and Information Technology of Shenzhen (201901161512) and "2011 Program" Peking-Tsinghua-IOP Collaborative Innovation Center of Quantum Matter. We acknowledge Electron Microscopy Laboratory in Peking University for the use of electron microscopes and financial support. We thank Dr. Chenglong Shi, Dr. Tracy Lovejoy and Dr. Jinlong Du for assistance in microscope operation.

**Conflict of Interest**

The authors declare no conflict of interest.

**Keywords**

hyperbolic phonon polaritons, α-MoO$_3$, electron energy loss spectroscopy

# REFERENCES

[1] J. D. Caldwell, L. Lindsay, V. Giannini, I. Vurgaftman, T. L. Reinecke, S. A. Maier, O. J. Glembocki, Nanophotonics **2015**, 4, 44.
[2] D. N. Basov, M. M. Fogler, F. J. Garcia de Abajo, Science **2016**, 354, aag1992.
[3] T. Low, A. Chaves, J. D. Caldwell, A. Kumar, N. X. Fang, P. Avouris, T. F. Heinz, F. Guinea, L. Martin-Moreno, F. Koppens, Nat Mater **2017**, 16, 182.

[4] Z. Jacob, Nat Mater **2014**, 13, 1081.

[5] E. Yoxall, M. Schnell, A. Y. Nikitin, O. Txoperena, A. Woessner, M. B. Lundeberg, F. Casanova, L. E. Hueso, F. H. L. Koppens, R. Hillenbrand, Nature Photonics **2015**, 9, 674.

[6] P. Li, M. Lewin, A. V. Kretinin, J. D. Caldwell, K. S. Novoselov, T. Taniguchi, K. Watanabe, F. Gaussmann, T. Taubner, Nat Commun **2015**, 6, 7507.

[7] A. J. Giles, S. Dai, I. Vurgaftman, T. Hoffman, S. Liu, L. Lindsay, C. T. Ellis, N. Assefa, I. Chatzakis, T. L. Reinecke, J. G. Tischler, M. M. Fogler, J. H. Edgar, D. N. Basov, J. D. Caldwell, Nat Mater **2018**, 17, 134.

[8] Z. Zheng, J. Chen, Y. Wang, X. Wang, X. Chen, P. Liu, J. Xu, W. Xie, H. Chen, S. Deng, N. Xu, Adv Mater **2018**, 30, e1705318.

[9] W. Ma, P. Alonso-Gonzalez, S. Li, A. Y. Nikitin, J. Yuan, J. Martin-Sanchez, J. Taboada-Gutierrez, I. Amenabar, P. Li, S. Velez, C. Tollan, Z. Dai, Y. Zhang, S. Sriram, K. Kalantar-Zadeh, S. T. Lee, R. Hillenbrand, Q. Bao, Nature **2018**, 562, 557.

[10] Z. Zheng, N. Xu, S. L. Oscurato, M. Tamagnone, F. Sun, Y. Jiang, Y. Ke, J. Chen, W. Huang, W. L. Wilson, A. Ambrosio, S. Deng, H. Chen, Science Advances **2019**, 5, eaav8690.

[11] G. Hu, Q. Ou, G. Si, Y. Wu, J. Wu, Z. Dai, A. Krasnok, Y. Mazor, Q. Zhang, Q. Bao, C. W. Qiu, A. Alu, Nature **2020**, 582, 209.

[12] G. Alvarez-Perez, T. G. Folland, I. Errea, J. Taboada-Gutierrez, J. Duan, J. Martin-Sanchez, A. I. F. Tresguerres-Mata, J. R. Matson, A. Bylinkin, M. He, W. Ma, Q. Bao, J. I. Martin, J. D. Caldwell, A. Y. Nikitin, P. Alonso-Gonzalez, Adv Mater **2020**, e1908176.

[13] O. Khatib, H. A. Bechtel, M. C. Martin, M. B. Raschke, G. L. Carr, ACS Photonics **2018**, 5, 2773.

[14] T. G. Folland, L. Nordin, D. Wasserman, J. D. Caldwell, Journal of Applied Physics **2019**, 125, 191102.

[15] R. Qi, R. Wang, Y. Li, Y. Sun, S. Chen, B. Han, N. Li, Q. Zhang, X. Liu, D. Yu, P. Gao, Nano Lett **2019**, 19, 5070.

[16] M. J. Lagos, A. Trugler, U. Hohenester, P. E. Batson, Nature **2017**, 543, 529.

[17] H. Lourenço-Martins, M. Kociak, Physical Review X **2017**, 7, 041059.

[18] O. L. Krivanek, T. C. Lovejoy, N. Dellby, T. Aoki, R. W. Carpenter, P. Rez, E. Soignard, J. Zhu, P. E. Batson, M. J. Lagos, R. F. Egerton, P. A. Crozier, Nature **2014**, 514, 209.

[19] A. A. Govyadinov, A. Konecna, A. Chuvilin, S. Velez, I. Dolado, A. Y. Nikitin, S. Lopatin, F. Casanova, L. E. Hueso, J. Aizpurua, R. Hillenbrand, Nat Commun **2017**, 8, 95.

[20] L. J. Allen, H. G. Brown, S. D. Findlay, B. D. Forbes, Microscopy (Oxf) **2018**, 67, i24.

[21] O. L. Krivanek, N. Dellby, J. A. Hachtel, J. C. Idrobo, M. T. Hotz, B. Plotkin-Swing, N. J. Bacon, A. L. Bleloch, G. J. Corbin, M. V. Hoffman, C. E. Meyer, T. C. Lovejoy, Ultramicroscopy **2019**, 203, 60.

[22] Y. Li, R. Qi, R. Shi, N. Li, P. Gao, Science Bulletin **2020**, 65, 820.

[23] A. Polman, M. Kociak, F. J. Garcia de Abajo, Nat Mater **2019**, 18, 1158.

[24] U. Hohenester, A. Trügler, P. E. Batson, M. J. Lagos, Physical Review B **2018**, 97, 165418.

[25] K. Eda, JOURNAL OF SOLID STATE CHEMISTRY **1991**, 95, 64.

[26] A. Losquin, M. Kociak, ACS Photonics **2015**, 2, 1619.

[27] L. Cheng, M. Shao, X. Wang, H. Hu, Chemistry **2009**, 15, 2310.

[28] Y. Cao, A. Manjavacas, N. Large, P. Nordlander, ACS Photonics **2015**, 2, 369.

[29] U. Hohenester, Computer Physics Communications **2014**, 185, 1177.

[30] T. Wang, P. Li, D. N. Chigrin, A. J. Giles, F. J. Bezares, O. J. Glembocki, J. D. Caldwell, T. Taubner, ACS Photonics **2017**, 4, 1753.

[31] J. S. Gomez-Diaz, A. Alù, ACS Photonics **2016**, 3, 2211.

[32] P. Li, I. Dolado, F. J. Alfaro-Mozaz, F. Casanova, L. E. Hueso, S. Liu, J. H. Edgar, A. Y. Nikitin, S. Vélez,

R. Hillenbrand, Science **2018**, 359, 892.

[33] J. H. Kang, S. Wang, Z. Shi, W. Zhao, E. Yablonovitch, F. Wang, Nano Lett **2017**, 17, 1768.

[34] A. Y. Nikitin, T. Low, L. Martin-Moreno, Physical Review B **2014**, 90, 041407.

[35] A. Ambrosio, M. Tamagnone, K. Chaudhary, L. A. Jauregui, P. Kim, W. L. Wilson, F. Capasso, Light Sci Appl **2018**, 7, 27.

[36] J. Taboada-Gutierrez, G. Alvarez-Perez, J. Duan, W. Ma, K. Crowley, I. Prieto, A. Bylinkin, M. Autore, H. Volkova, K. Kimura, T. Kimura, M. H. Berger, S. Li, Q. Bao, X. P. A. Gao, I. Errea, A. Y. Nikitin, R. Hillenbrand, J. Martin-Sanchez, P. Alonso-Gonzalez, Nat Mater **2020**, (published online, DOI: 10.1038/s41563-020-0665-0).

[37] J. Zhou, Y. Yang, Y. Yang, D. S. Kim, A. Yuan, X. Tian, C. Ophus, F. Sun, A. K. Schmid, M. Nathanson, H. Heinz, Q. An, H. Zeng, P. Ercius, J. Miao, Nature **2019**, 570, 500.

[38] K. Dabov, A. Foi, V. Katkovnik, K. Egiazarian, IEEE Transactions on Image Processing **2007**, 16, 2080.

[39] W. H. Richardson, J. Opt. Soc. Am. **1972**, 62, 55.

[40] L. B. Lucy, The Astronomical Journal **1974**, 79, 745.

[41] U. Hohenester, A. Trügler, Computer Physics Communications **2012**, 183, 370.

[42] J. Waxenegger, A. Trügler, U. Hohenester, Computer Physics Communications **2015**, 193, 138.